\documentclass[prd,preprint,eqsecnum,amsmath,amssymb,nofootinbib]{revtex4}
\usepackage{graphicx, tikz}
\usepackage{caption, subcaption}
\usepackage{amssymb,amsmath, mathrsfs}
\usepackage{amssymb, graphics, setspace}
\usepackage{hyperref}
\usepackage[all]{xy}
\newcommand{\be}{\begin{equation}}
\newcommand{\ee}{\end{equation}}
\newcommand{\bea}{\begin{eqnarray}}
\newcommand{\eea}{\end{eqnarray}}
\newcommand{\bes}{\begin{subequations}}
\newcommand{\ees}{\end{subequations}}

\begin{document}

\title{Quantum energy momentum tensor and equal time correlations in a Reissner-Nordstr\"om black hole}
\author{Roberto~Balbinot}
\email{roberto.balbinot@unibo.it}
\affiliation{Dipartimento di Fisica dell'Universit\`a di Bologna and INFN sezione di Bologna, Via Irnerio 46, 40126 Bologna, Italy
%
}
\author{Alessandro~Fabbri}
\email{afabbri@ific.uv.es}
\affiliation{Departamento de F\'isica Te\'orica and IFIC, Universidad de Valencia-CSIC, C. Dr. Moliner 50, 46100 Burjassot, Spain
}

\bigskip\bigskip

\begin{abstract}
We consider a Reissner-Nordstr\"om black hole formed by the collapse of a charged null shell. The renormalised expectation values of the energy momentum tensor operator for a massless scalar field propagating in the
2D section of this spacetime are given. We then analyse the across the horizon correlations of the related energy density operator for free falling observers to reveal the correlations  between the Hawking particles and their interior partners.
\end{abstract}
\maketitle
\section{Introduction}

The existence of quantum correlations across the horizon in black holes associated to Hawking radiation \cite{hawking} has attracted increasing interest especially in the community of analogue models \cite{Barcelo:2005fc}. As it is well known the Hawking effect consists in the conversion of quantum vacuum fluctuations in pairs of on shell particles \cite{Brout:1995rd} (phonons in the case of acoustic black holes). A member of the pair (the Hawking particle) carries positive Killing energy and emerges outside the horizon and propagates to infinity constituting the asymptotic thermal flux. The other member of the pair (the “partner”), has negative Killing energy and is created inside the horizon 
of the black hole (hereafter BH).

\begin{figure}[h]
\centering \includegraphics[angle=0, height=3.in] {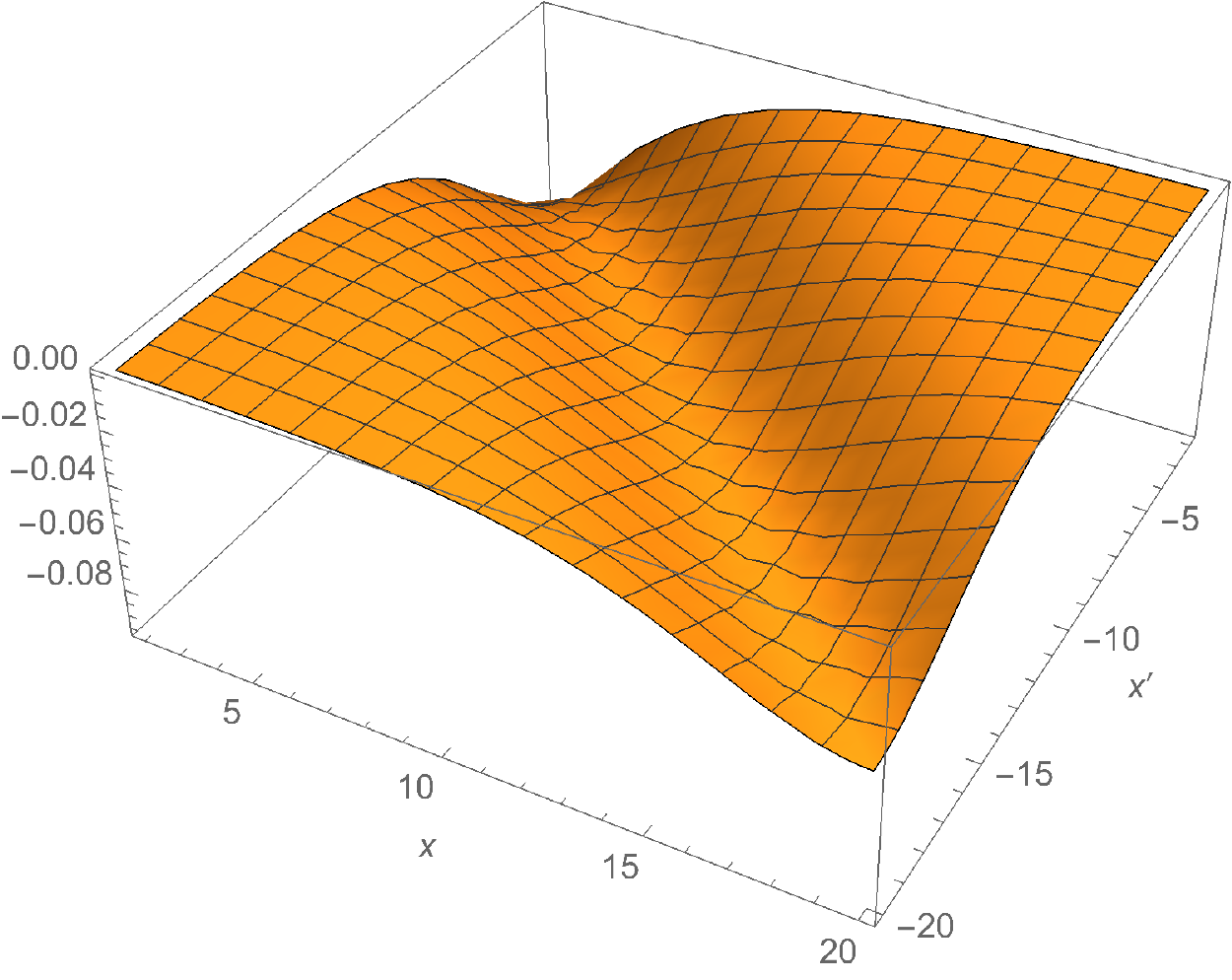}
\caption{Theoretically predicted  equal-time density-density correlation function from the model in Refs. \cite{Fabbri:2020unn, Balbinot:2021bnp}, $x'$ is inside the horizon and $x$ is outside. The peak is at $x=-x'$.}
\label{figuno}
\end{figure}

The correlations between the Hawking particle and its partner is a distinctive feature of the Hawking effect which should manifest itself in the appearance of a characteristic peak \cite{Balbinot:2007de,Fabbri:2020unn, Balbinot:2021bnp} in the equal-time correlation functions across the horizon, see Fig. (\ref{figuno}).
 This peak has indeed been observed, see for instance Fig. (3a) of Ref. \cite{jeff2}, in analogue black holes (BHs) formed by a Bose-Einstein condensate (BEC) undergoing a transition from a subsonic flow to a supersonic one \cite{jeff1, jeff2, jeff3}.  This remarkable fact represents up to today the best experimental evidence of the existence of Hawking radiation.
 
 One should remark that the features of this correlation function highly depend on the fate of the partner and hence on the spacetime structure inside the horizon. 

\begin{figure}[h]
\centering \includegraphics[angle=0, height=3.in] {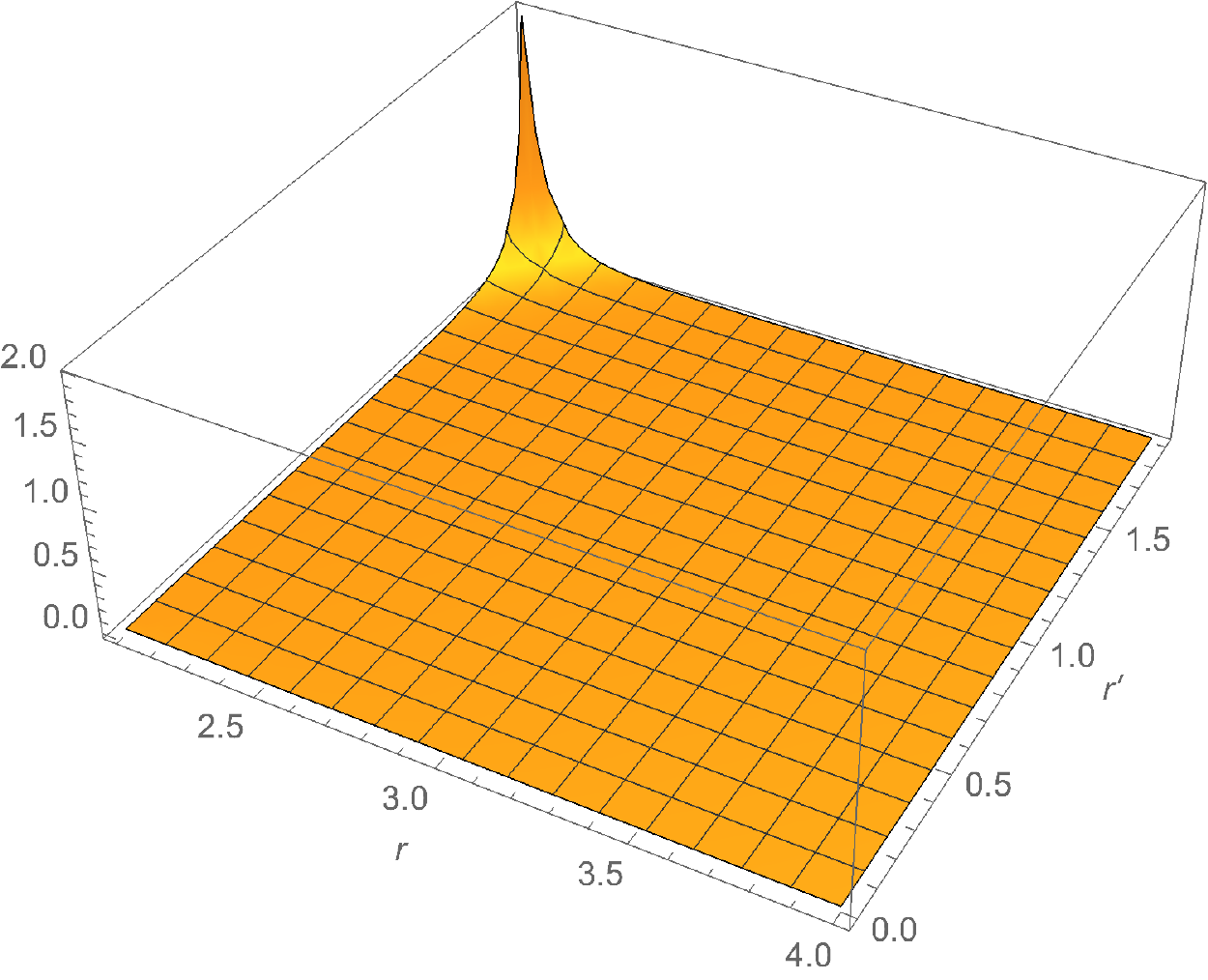}
\caption{Density-density correlation in the Schwarzschild black hole, see Ref. \cite{Balbinot:2021bnp} for details, $r>2m$ and $r'<2m$ ($m=1$).}
\label{figdue}
\end{figure}

Note that in the acoustic BHs so far realized in laboratory the supersonic region inside the horizon does not end at a singularity, as it happens instead in the gravitational case, but continues asymptotically towards a homogeneous configuration reached eventually by the partners. The presence of a central singularity, mimicked in the acoustic case by a sink absorbing both the condensate atoms and the phonons\footnote{Experimental setups realizing these configurations are in construction. We thank I. Carusotto for this information.}, has a dramatic effect on the equal time correlation functions: the peak does not appear, see Fig. (\ref{figdue}).
This because the Hawking particles and their partners are created in a region of extension of the order of $\frac{1}{\kappa}$, where $\kappa$ is the surface gravity of the BH, across the horizon, the so called “quantum atmosphere” \cite{qatm1,qatm3}. And when the Hawking particle emerges out of vacuum fluctuations from the quantum atmosphere, the corresponding partner has already been swallowed by the central singularity and no sign of correlations appears: they are lost in the singularity. To observe them one has to consider correlations at unequal equal times in order to catch the partner before it disappears.
These aspects have emerged in a recent study of the quantum correlations across the horizon in a Schwarzschild BH \cite{Balbinot:2021bnp}. Given this strong dependence on the inner metric of the BH, here we extend the analysis to the case of a Reissner Nordstr\"om (RN) BH whose internal structure, because of the presence of an inner horizon, is much more intriguing.

The plan of the paper is the following. In section II we present a simple model for the formation of a RN BH by the collapse of a null charged shell triggering the Hawking effect. In section III, in preparation to the study of the quantum correlations across the horizon, we discuss the renormalized energy momentum tensor associated to a massless scalar quantum field propagating in the two dimensional section of the collapse spacetime of our model. We will concentrate on the expectation value of energy density operator as measured by a free falling observer with particular attention to what happens close to the inner horizon. In section IV we analyze the across horizon correlation functions of the above energy density operator. Finally section V contains our conclusions.

\section{Forming a Reissner-Nordstr\"om BH}
\label{s2}

It is well known that the formation of a BH triggers a vacuum instability leading to the emission of thermal radiation far away from the BH horizon. We will use a simple model for the formation of a Reissner-Nordstr\"om BH, namely the collapse of an ingoing charged null shell located at $v=v_0$, where $v$ is an ingoing null coordinate. 

The metric of the spacetime is of a Vaidya form
\be \label{dueuno} ds^2=-(1-\frac{2m(r)}{r})dv^2+2dvdr +r^2 d\Omega^2 \ , \ee
where $d\Omega^2=d\theta^2+\sin^2\theta d\varphi^2$ and $m(r)$ is such that
\bea m(r) &=& 0\ , \ \ \ \ \ \ \ \ \ \ \ v<v_0 \label{dueduea}\ ,  \\
m(r) &=& m-\frac{Q^2}{r}\ , \ \ v>v_0 \label{duedueb} \ . \eea

For $v<v_0$ the spacetime is Minkowski one and we can write the metric in a double null form 
\be \label{duetre} ds^2=-du_{in}dv +r^2(u_{in},v)d\Omega^2 \ , \ee
where 
\be \label{duequattro} r_{in}= \frac{v-u_{in}}{2} \ee
and the double null coordinates are
\be \label{duecinque} u_{in}=t_{in}-r\ , \ \ v=t_{in}+r \ . \ee

In the future of the shell (i.e. $v>v_0$) the metric is the RN one which can also be given in a double null form 
\be ds^2=-f(r)dudv +r^2d\Omega^2 \ , \label{duesei} \ee 
\be  f(r)=1-\frac{2m}{r}+\frac{Q^2}{r^2} \ , \label{duesette} \ee
where 
\be \label{dueotto} u=t-r^* \ , v=t+r^*\ , \ee
and $r^*$ is the Regge-Wheeler tortoise coordinate
\be \label{duenove} r^*= \int \frac{dr}{f}=r +\frac{1}{2\kappa_+}\ln|\kappa_-(r-r_+)|-\frac{1}{2\kappa_-}\ln|\kappa_-(r-r_-)|\ ,\ee
where 
\be \label{duedieci} r_{\pm}=m\pm\sqrt{m^2-Q^2} \ee
are the two horizons and 
\be \kappa_\pm=\frac{|f'(r)|_{r_{\pm}}}{2}=\frac{r_+-r_-}{2r_{\pm}^2}=\frac{\sqrt{m^2-Q^2}}{r_\pm^2}\  \label{dueundici} \ee
the corresponding surface gravities. 

\begin{figure}[h]
\centering \includegraphics[angle=0, height=3.5in] {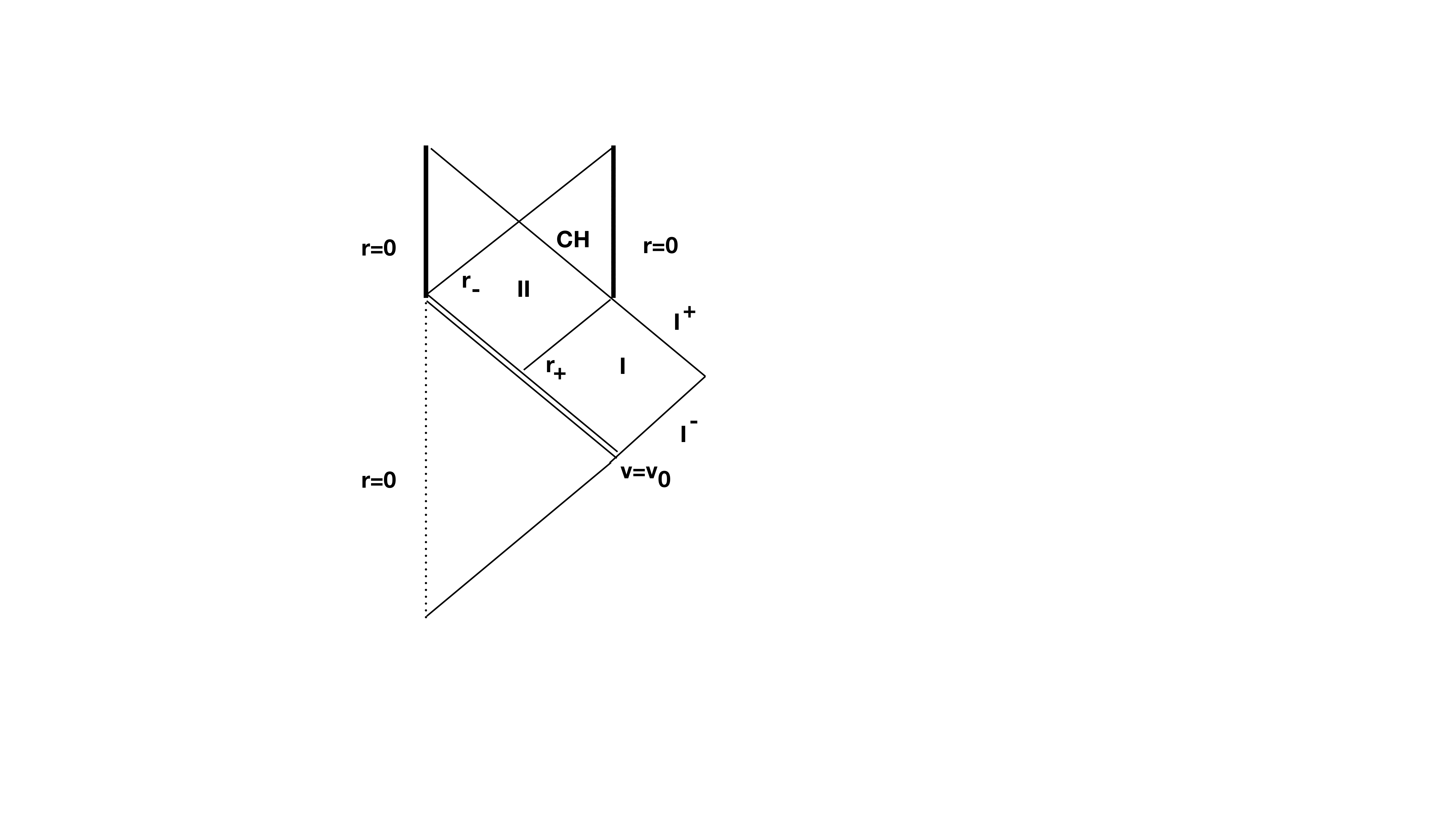}
\caption{Penrose diagram of the spacetime. The regions considered in the paper are the asymptotic one (I) and the one between the horizons (II).}
\label{figtre}
\end{figure}

In order for eq. (\ref{duesei}) to describe a BH we require $m^2>Q^2$. We will refer to $r_+$ (where $u=+\infty$) simply as the event horizon of the BH, while we call the ``inner horizon" the outgoing sheet of $r_-$ (where $u=-\infty$) and the ``Cauchy horizon" the ingoing one (where $v=+\infty$). The relevant Penrose diagram is given in Fig. (\ref{figtre}).

\begin{figure}[h]
\centering \includegraphics[angle=0, height=3.5in] {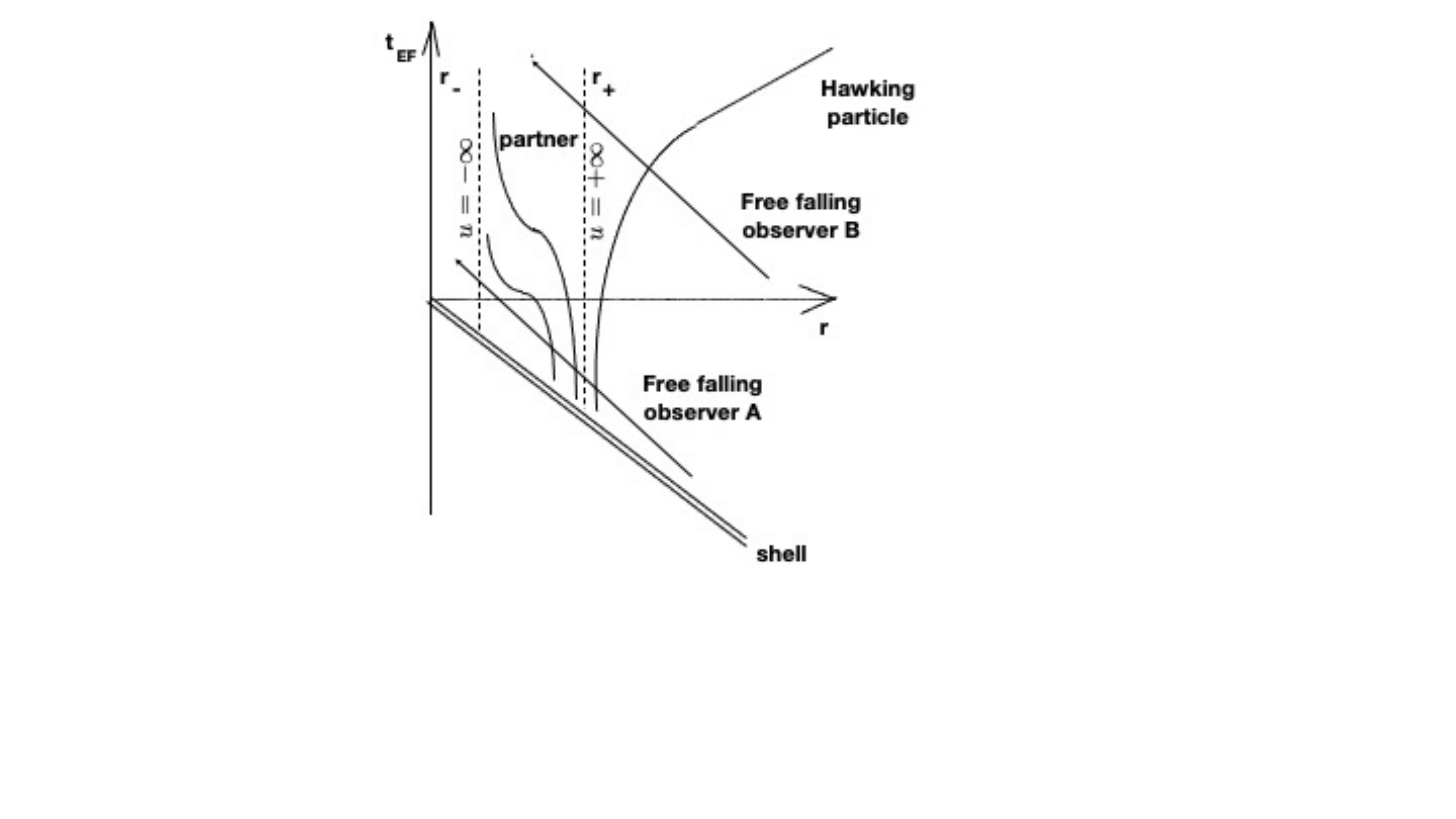}
\caption{Spacetime diagram in the RN portion of the spacetime representing three $u=cst$ curves, one for the Hawking particle, one for the partner (both at large positive $u$) and one for an intermediate value of $u$, and the trajectories of two free-falling observers crossing $r_+$ and $r_-$ at, respectively, early (observer $A$) and late time (observer $B$) after BH formation.}
\label{figquattro}
\end{figure}

We introduce in regions I and II a time coordinate (Eddington-Finkelstein time) defined as 
\be \label{tef} t_{EF}=v-r\ ,\ee and
\bea 
v &=& t_{EF}+r\ , \label{duequindicia} \\
u &=& t_{EF} + r-2r^*=t_{EF}-r-\frac{1}{\kappa_+}\ln|\kappa_+(r-r_+)| +\frac{1}{\kappa_-}\ln|\kappa_-(r-r_-)| \label{duequindicib} \eea
are RN null coordinates. Note that at the Cauchy horizon $t_{EF}$ goes to $+\infty$. A spacetime diagram representing three characteristic curves $u=cst$ in the RN portion of the spacetime is given in Fig. (\ref{figquattro}). 
The one denoted Hawking particle starts close outside the event horizon $r_+$, while the one denoted partner starts just inside $r_+$; both of them are characterized by a large positive $u$. Another $u=cst$ curve at intermediate $u$ is also represented. Note the characteristic piling up of the $u=cst$ trajectories along $r_-$. Finally we have also depicted the trajectories of two free falling observers crossing the event (and the inner) horizon at short time after the formation of the BH (observer A) and at late-time (observer B).

Matching the two line-elements eq. (\ref{duetre}) and (\ref{duesei}) along the shell $v=v_0$ we get \cite{Balbinot:2021bnp}
\be \label{duesedici} u=u_{in}-\frac{1}{\kappa_+}\ln | \kappa_+(v_0-u_{in}-2r_+)| +\frac{1}{\kappa_-}\ln |\kappa_- (v_0-u_{in}-2r_-)| \ . \ee
Using this we can extend the $u_{in}$ coordinate in the RN portion of the spacetime. 

The event horizon ($r=r_+,\ u=+\infty$) corresponds to 
\be u_{in}|_{eh}=v_0-2r_+=0 \ . \label{duediciassette} \ee
To simplify the notation we have set $v_0=2r_+$ so that the event horizon ($u=+\infty$) corresponds to $u_{in}=0$. The inner horizon ($u=-\infty$) corresponds to 
\be \label{duediciotto} u_{in}|_{ih}=2(r_+-r_-)\ .\ee 
Eq. (\ref{duesedici}) cannot be inverted analytically, however we can get the following limiting behaviours near the event horizon (large positive $u$)
\be u\simeq -\frac{1}{\kappa_+}\ln |-\kappa_+ u_{in}| \label{duediciannovea} \ee
and near the inner horizon (large negative $u$)
\be \label{duediciannoveb} u\simeq \frac{1}{\kappa_-}\ln|\kappa_-(2(r_+-r_-)-u_{in})| \ee
which will be useful in the rest of the paper. Note from these that $u_{in}$ behaves in the limits considered as the Kruskal coordinate of the corresponding horizon. 

\section{Quantum Field Theory in the 2D RN spacetime}
\label{s3}

We consider now a massless scalar quantum field $\hat\phi$ propagating in the 2D section ($\theta=\varphi=cst$) of our collapse spacetime. Its equation of motion is 
\be \label{treuno} \hat\Box\hat \phi =0\ , \ee where
$\hat\Box=\nabla_\mu\nabla^\mu$. The field is supposed to be in a quantum state called $|in\rangle$, which corresponds to Minkowski vacuum on past null infinity (see Fig. (\ref{figtre})), which is a Cauchy surface for our field equation.

Its energy-momentum  tensor operator is
\be \label{tredue} \hat T_{ab}(\hat\phi)= \partial_a\hat\phi\partial_b\hat\phi-\frac{g_{ab}}{2}\partial^c\hat\phi\partial_c\hat\phi \ee
and the corresponding renormalized expectation values in the $|in\rangle$ state are the following
(see \cite{Balbinot:2007kr} for details). 
For $v<v_0$ 
\be \label{tretre} \langle in|\hat T_{ab} |in\rangle =0\ . \ee
For $v>v_0$ 
\bea \label{trequattroa} \langle in|\hat T_{vv}|in\rangle &=& -\frac{1}{192\pi}\left( f'(r)^2-2f(r)f''(r) \right) \\ &=&
\frac{1}{24\pi}\left( -\frac{m}{r^3}+\frac{3}{2} \frac{(m^2+Q^2)}{r^4}-\frac{3mQ^2}{r^5}+\frac{Q^4}{r^6}\right) = \langle B|\hat T_{vv}|B\rangle \ , \nonumber 
\eea
where a prime ``\ $'$ " indicates derivative with respect to $r$, 
\be \label{trequattrob}  \langle in|\hat T_{uu}|in\rangle =  \langle B|\hat T_{uu}|B\rangle 
-\frac{1}{24\pi}\{ u_{in},u\} \ee
with $\langle B|\hat T_{uu}|B\rangle =\langle B|\hat T_{vv}|B\rangle$ and $\{\ ,\ \}$ is the Schwarzian derivative   calculated from eq. (\ref{duesedici}). Explicitly,
\be \label{trecinque}
\{ u_{in}, u \} = \frac{3}{2}\kappa_+^2 \frac{\left( 1- \frac{\kappa_+}{\kappa_-} \frac{u_{in}^2}{(u_{in}-2(r_+-r_-))^2}\right)^2}{\left(1-\kappa_+ u_{in}-\frac{\kappa_+}{\kappa_-}\frac{u_{in}}{(u_{in}-2(r_+-r_-))}\right)^4}-2\kappa_+^2\frac{\left( 1- \frac{\kappa_+}{\kappa_-} \frac{u_{in}^3}{(u_{in}-2(r_+-r_-))^3}\right)^2}{\left(1-\kappa_+ u_{in}-\frac{\kappa_+}{\kappa_-}\frac{u_{in}}{(u_{in}-2(r_+-r_-))}\right)^3}\ . \ee 
Finally,
\be \label{tresei} 
\langle in|\hat T_{uv}|in\rangle = \langle B|\hat T_{uv}|B\rangle=-\frac{1}{24\pi}(1-\frac{2m}{r}+\frac{Q^2}{r^2})(\frac{m}{r^3}-\frac{3}{2}\frac{Q^2}{r^4})\ . \ee
Note that, because of conformal invariance, $\hat T_{ab}$ is traceless, i.e. $\hat T_{uv}=0$. The nonvanishing result of eq. (\ref{tresei}) comes from the conformal anomaly, which in this case is simply proportional to the Ricci scalar.
In the r.h.s. of eqs. (\ref{trequattroa}), (\ref{trequattrob}), (\ref{tresei}) we have indicated the expectation values calculated in the Boulware vacuum $|B\rangle $, which describes the local vacuum polarization associated to the spacetime curvature. Note that at the horizons 
\be \label{tresette} 
\langle B|\hat T_{uu}|B\rangle |_{r_{\pm}}=\langle B|\hat T_{vv}|B\rangle |_{r_{\pm}}=-\frac{\kappa_{\pm}^2}{48\pi} \ . \ee
The Schwarzian derivative in eq. (\ref{trequattrob}) is associatd to the particles creation induced by the formation of the BH. These propagate along $u=cst$ trajectories. Using the form of eq. (\ref{duediciannovea}) and (\ref{duediciannoveb}) the following asymptotic behaviours of the Schwarzian derivative term can be found for $u\to +\infty$ ($u_{in}\to 0$ in eq. (\ref{trecinque}))
\be \label{treotto} -\frac{1}{24\pi} \{ u_{in}, u \}=\frac{1}{48\pi} \kappa_+^2\ , \ee
and
\be \label{trenove} -\frac{1}{24\pi} \{ u_{in}, u \}=\frac{1}{48\pi} \kappa_-^2\  \ee
for $u\to -\infty$ ($u_{in}\to 2(r_+-r_-)$ in eq. (\ref{trecinque})) .
A free falling observer measures the following energy density associated to the field (see also \cite{Balbinot:2021bnp}) 
\be \label{tredieci} 
\rho=\langle in|\hat\rho| in\rangle=\langle in | \hat T_{ab} u^au^b |in\rangle \ee
where $u^a$ is the four velocity vector of the observer trajectory. One easily obtains

 \be \hat\rho =  
  \frac{ \left(E+\sqrt{E^2-f}\right)^2}{f^2}\langle in| \hat T_{uu}| in\rangle  +  \frac{ \left(E-\sqrt{E^2-f}\right)^2}{f^2}\langle in | \hat T_{vv} | in\rangle + \frac{2}{f} \langle in | \hat T_{vv} | in\rangle \ ,\label{treundici}  \ee 
where $E$ is the conserved Killing energy of the observer. For the moment we set $E=1$, which corresponds to a geodesic starting with zero velocity at infinity. 

An exact analytical expression of $\rho$ as a function of $r$ and $t_{EF}$ along the trajectory of the observer cannot be given since we are unable to invert the relation (\ref{duesedici}) allowing to express the Schwarzian derivative in terms of the above coordinates. We can however deduce some limiting behaviour of $\rho$. 

For an observer at infinity ($r\to +\infty$) the vacuum polarization terms vanish and if we consider the observer at late time ($u\to +\infty$) we can use eq. (\ref{treotto}) and find
\be \label{tredodici} \rho = \frac{\kappa_+^2}{48\pi} \ , \ee
which corrresponds to a thermal flux of massless particles at the Hawking temperature 
\be \label{tretredici} T_H=\frac{\hbar \kappa_+}{2\pi}\ . \ee
Now consider observers as they cross the event horizon $r=r_+$. 
First, let us rewrite eq. (\ref{treundici}) (for $E=+1$) in the form
\be \label{trequattordici}
\rho =  \frac{ \left(1+\sqrt{\frac{r_++r_-}{r}-\frac{r_+r_-}{r^2}} \right)^2}{f^2}\langle in| \hat T_{uu}| in\rangle  + \frac{ \langle in | \hat T_{vv} | in\rangle}{\left(1+\sqrt{\frac{r_++r_-}{r}-\frac{r_+r_-}{r^2}} \right)^2 } + \frac{2}{f} \langle in | \hat T_{uv} | in\rangle \  . \ee 
To evaluate the Schwarzian derivative term entering $\langle in| \hat T_{uu}| in\rangle$ one should note that the event horizon $r_+$ corresponds to $u=+\infty$ and in that limit, using again (\ref{treotto}), we can write 
\be \label{trequindici}
\langle in| \hat T_{uu}| in\rangle = \langle B| \hat T_{uu}| B\rangle +\frac{\kappa_+^2}{48\pi}
=\frac{1}{48\pi} \frac{(r-r_+)^2}{r^2}\left( \kappa_+^2(1+\frac{2r_+}{r}+\frac{3r_+^2}{r^2})+\frac{(\frac{r_-^2}{r_+}-3r_-)}{r^3}+\frac{2r_-^2}{r^4}\right)\ ,  \ee
which is exactly the value one would obtain in the Unruh vacuum associated to the event horizon. 
We see that it vanishes at $r_+$ making the first term in eq. (\ref{trequattordici}) finite as $r\to r_+$. The vacuum polarization piece, see eq. (\ref{tresette}), is exactly canceled by the Schwarzian derivative one (see eq. (\ref{treotto})). The last term in eq. (\ref{trequattordici}) is also regular (see eq. (\ref{tresei})) and putting everything together we can write
\bea \label{tresedici}  \rho =  && \frac{1}{48\pi}\Big[ \frac{ \left(1+\sqrt{\frac{r_++r_-}{r}-\frac{r_+r_-}{r^2}} \right)^2 r^2}{(r-r_-)^2}\left( \kappa_+^2(1+\frac{2r_+}{r}+\frac{3r_+^2}{r^2})+\frac{(\frac{r_-^2}{r_+}-3r_-)}{r^3}+\frac{2r_-^2}{r^4}\right) + \ \ \ \ \ \ \  \nonumber  \\ && \frac{-\frac{(r_++r_-)}{r^3} +\frac{\frac{3}{4}(r_++r_-)^2+3r_+r_-}{r^4}-\frac{3r_+r_-(r_++r_-)}{r^5}+\frac{2r_+^2r_-^2}{r^6}}{\left(1+\sqrt{\frac{r_++r_-}{r}-\frac{r_+r_-}{r^2}} \right)^2}  -\frac{2(r_++r_-)}{r^3}+\frac{6r_+r_-}{r^4}  \Big]  \eea
giving the value 
\be \label{trediciassette} \rho|_{eh} = \frac{1}{48\pi} \left( \frac{6}{r_+^2} -\frac{6r_-}{\kappa_+ r_+^4}-\frac{\kappa_+^2}{4} -\frac{2(r_+-2r_-)}{r_+^3} \right) \ee
on the event horizon. 

The behaviour of $\rho$ as the observers approach the inner horizon $r_-$ is more delicate. If the observers cross the event horizon $u=+\infty$ (and hence the inner horizon) at early time after the formation of the BH (see trajectory $A$ in Fig. (\ref{figquattro})) they leave very rapidly the large (positive) $u$ region and the measured energy density smoothly approaches the limiting value
\be \label{trediciotto} \rho|_{ih} = \frac{1}{48\pi} \left( \frac{6}{r_-^2} -\frac{2}{\kappa_- r_+r_-^2}-\frac{\kappa_-^2}{4} -\frac{2(r_--2r_+)}{r_-^3} \right)\ . \ee
Again one sees the cancellation between the vacuum polarization  (\ref{tresette}) with (\ref{trenove}). 

However, if the observer is approaching the inner horizon at very late time (trajectory $B$ in Fig (\ref{figquattro})) things are strikingly different \cite{Jacobson:1997ge}. The outgoing null rays inside $r_+$ peel away from the event horizon and asymptotically approach the inner horizon $r_-$ (see Fig. (\ref{figquattro})) in a diverging $t_{EF}$. So the partners emitted just outside $r_+$ with $u\to +\infty$ get in the above limit arbitrarily close to the inner horizon where $f\ll 1$. So at very late time our free falling observers meet all of these partners in a region where $f\ll 1$ but $u$ is still very large and positive. 
Using eq. (\ref{treotto}) for the Schwarzian derivative we have that 
\be \label{treventi} \langle in| \hat T_{uu}| in\rangle = \langle B| \hat T_{uu}| B\rangle +\frac{\kappa_+^2}{48\pi}\ . \ee
So the energy density he measures, in virtue of eq. (\ref{tresette}) evaluated at $r_-$, grows as
$\frac{\kappa_+^2-\kappa_-^2}{f^2}$ coming from the first term in eq. (\ref{trequattordici}); the other two are bounded. Being $\kappa_- > \kappa_+$ this ever increasing density is negative. 

Finally, we can consider the case of an observer approaching the Cauchy horizon starting from inside the event horizon. For this observer $E$ is negative, say $E=-1$, and eq. (\ref{trequattordici}) is replaced by 
\be \label{treventuno}
\rho =   \frac{ \langle in | \hat T_{uu} | in\rangle}{\left(1+\sqrt{\frac{r_++r_-}{r}-\frac{r_+r_-}{r^2}} \right)^2 }+  \frac{ \left(1+\sqrt{\frac{r_++r_-}{r}-\frac{r_+r_-}{r^2}} \right)^2}{f^2}\langle in| \hat T_{vv}| in\rangle + \frac{2}{f} \langle in | \hat T_{uv} | in\rangle \   \ee 
showing that $\rho$ diverges as $-\frac{\kappa_-^2}{f^2}$ due to vacuum polarization. 

\section{Particle-partner correlations}
\label{s4}

The analysis of the expectation values of the energy momentum tensor operator in the last section has shown how vacuum polarization effects and particles creation mix up and, beside kinematical effects associated to the world line of the observer, both contribute to the measured energy density. Only asymptotically the energy momentum tensor describes an outgoing flux of particles at the Hawking temperature. On the other hand, inside the horizon, because of the non vanishing vacuum polarization, the tensor does not simply describes an ingoing flux of partners.
To reveal the genuine pairs creation process which is at the basis of the Hawking effect one should try to highlight the existing quantum correlations between the Hawking particles and their associated partners . To this end in this section we discuss the correlation functions
$G(x,x')$ of the energy density operator
\be \label{quattrouno} G(x,x')\equiv \langle in| \hat\rho(x)\hat\rho (x')|in\rangle\ , \ee
where 
 \be \label{quattrodue}  \hat\rho = 
  \frac{ \left(E+\sqrt{E^2-f}\right)^2}{f^2}\hat T_{uu} +  \frac{ \left(E-\sqrt{E^2-f}\right)^2}{f^2}\hat T_{vv}\   \ee  
  and $\hat T_{ab}$ is the energy-momentum tensor operator defined in eq. (\ref{tredue}). $E$ is the conserved Killing energy of the observer. 
 The starting point is the 2-point function for the $|in\rangle $ vacuum 
\be \label{quattrotre}
\langle in| \hat \phi(x)\hat  \phi(x')|in\rangle = -\frac{\hbar}{4\pi}\ln (u_{in}-u_{in}')(v-v')\ . \ee
Out of this one can build the fundamental object of our calculation
\be \label{quattroquattro} 
\partial_u\partial_{u'} \langle in|\hat\phi(x)\hat\phi(x')|in\rangle =-\frac{\hbar}{4\pi} \frac{du_{in}}{du}
 \frac{du_{in}'}{du'}  \frac{1}{(u_{in}-u_{in}')^2}\ . \ee
The $\hat T_{uu}$ correlator is then (see \cite{Balbinot:2021bnp} for details)
\be \label{quattrocinque} 
\langle in|\hat T_{uu}(x)|\hat T_{uu}(x')|in\rangle = \Big[ \partial_u\partial_{u'} \langle in|\hat\phi(x)\hat\phi(x')|in\rangle   \Big]^2 \ . \ee
This is the relevant one to discuss the Hawking-partner correlations. 

We can write the density correlator as follows (both observers have $E=+1$) 
\be \label{quattrosei} G(x,x')= \frac{ \langle in|\hat T_{uu}(x)\hat T_{u'u'}(x')|in\rangle}{\left(1-\sqrt{\frac{2m(r)}{r}} \right)^2\left(1-\sqrt{\frac{2m(r')}{r}} \right)^2}+  \frac{ \langle in|\hat T_{vv}(x)\hat T_{v'v'}(x')|in\rangle}{\left(1+\sqrt{\frac{2m(r)}{r}} \right)^2\left(1+\sqrt{\frac{2m(r')}{r}} \right)^2}\ . \ee
As said before, for our purposes we focus on the first term in eq. (\ref{quattrosei}), take the point $x$ outside the event event horizon $r_+$ and $x'$ in between the inner horizon and the event one. 

Consider first a trajectory of our inner observer like the one labelled $A$ in Fig. (\ref{figdue}) and take $x'$ inside and close to $r_+$. We can then use eq. (\ref{duediciannovea}), namely 
\be \label{quattrosette} u'\sim -\frac{1}{\kappa_+}\ln|-\kappa_+ u_{in}'|\ . \ee
Similarly, taking $x$ close outside $r_+$, using again eq. (\ref{duediciannovea}), 
\be \label{quattrootto} u \sim -\frac{1}{\kappa_+}\ln|-\kappa_+ u_{in}|\ . \ee
Inserting these in eq. (\ref{quattroquattro}) we have 
\be \label{quattronove} 
\frac{ \langle in|\hat T_{uu}(x)\hat T_{u'u'}(x')|in\rangle}{\left(1-\sqrt{\frac{2m(r)}{r}} \right)^2\left(1-\sqrt{\frac{2m(r')}{r}} \right)^2}
\sim \frac{1}{ \left(1-\sqrt{\frac{2m(r)}{r}} \right)^2\left(1-\sqrt{\frac{2m(r')}{r}} \right)^2}\left( \frac{\hbar \kappa_+^2}{16\pi\cosh^2(\frac{\kappa_+}{2}(u-u'))}\right)^2 . \ee

\begin{figure}[h]
\centering
\begin{subfigure}[h]{0.45\textwidth}
{\includegraphics[width=2in]{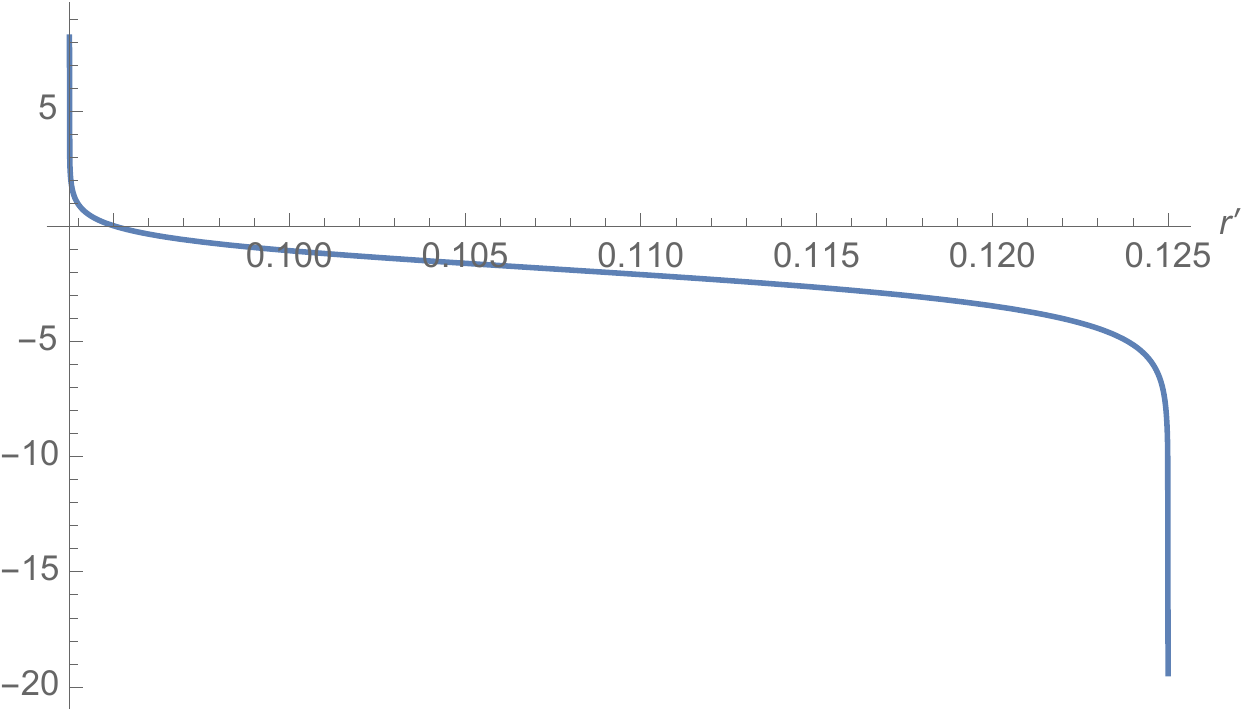}}
\caption{}
\label{5a}
\end{subfigure}
\begin{subfigure}[h]{0.45\textwidth}
{ \includegraphics[width=2in]{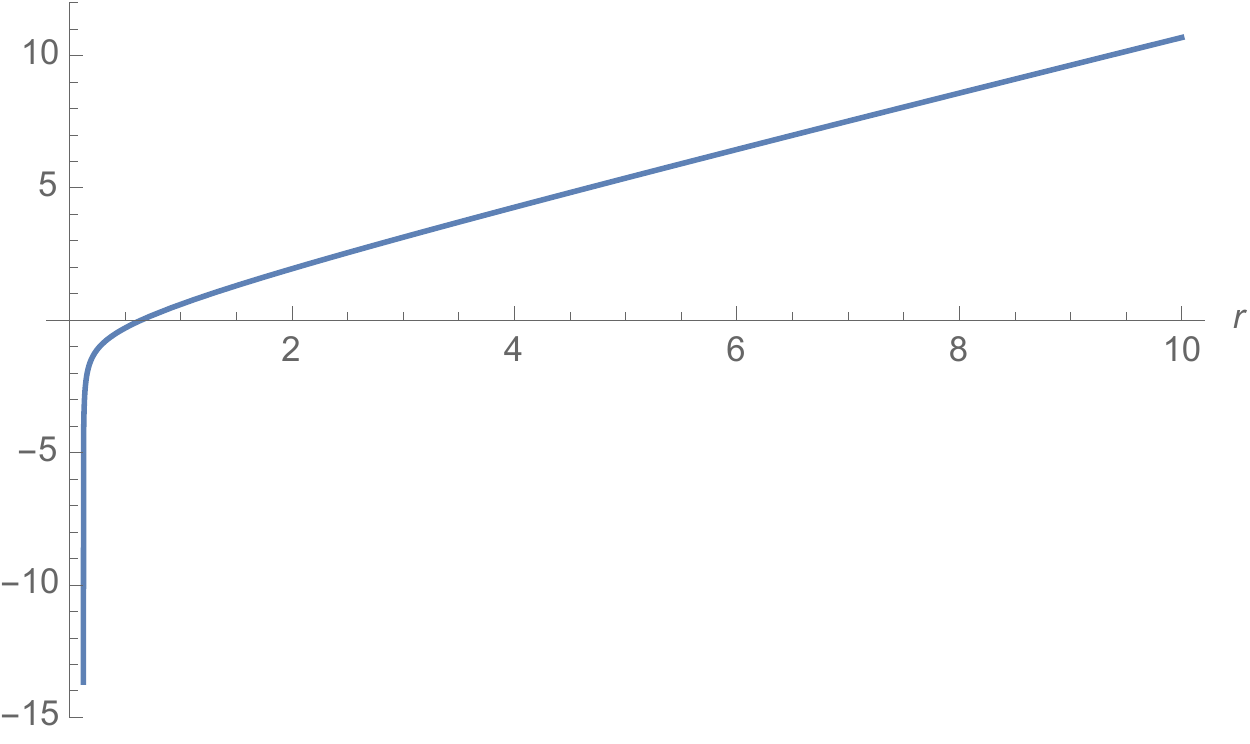}}
\caption{}
\label{5b}
\end{subfigure}
\caption{\label{figcinque} Plot of the lhs (a) and rhs (b)   of eq. (\ref{quattrodieci}). In this figure and in those that follow we consider  $r_+=\frac{1}{8}=0.125,\ r_-=\frac{3}{32}=0.09375,\ \kappa_+=1,\ \kappa_-=\frac{16}{9}$.  }
\end{figure}

One sees the appearance of the well known $\cosh^2$ term modulated by the geometrical prefactors containing $\sqrt{\frac{2m(r)}{r}}$. For an acoustic BH in a BEC the density-density correlator has a structure similar to eq. (\ref{quattronove}) (more precisely the square of it). In that case the points $x$ and $x'$ are typically taken far away on both sides of the acoustic (single) horizon where the medium is homogeneous and the geometric prefactors are just harmless constants, and the peak of the correlator corresponds to the maximum of the $\cosh^{-2}$ term, namely $u=u'$, i.e. along the trajectories of the Hawking particle and its corresponding partner. This condition at equal time in our case would be 
\be \label{quattrodieci}
 r'+\frac{1}{\kappa_+}\ln|\kappa_+(r'-r_+)|-\frac{1}{\kappa_-}\ln|\kappa_-(r'-r_+)|=
r+\frac{1}{\kappa_+}\ln|\kappa_+(r-r_+)|-\frac{1}{\kappa_-}\ln|\kappa_-(r-r_+)|
\ee


\noindent where remember that $r_-<r'<r_+$ and $r>r_+$. The plot of eq. (\ref{quattrodieci}) is given in Fig. (\ref{figcinque}): they both go to $+\infty$ for both $r\to r_-$ and $r\to +\infty$ and to $-\infty$ for 
$r\to r_+$.

\begin{figure}[h]
\centering \includegraphics[angle=0, height=2in] {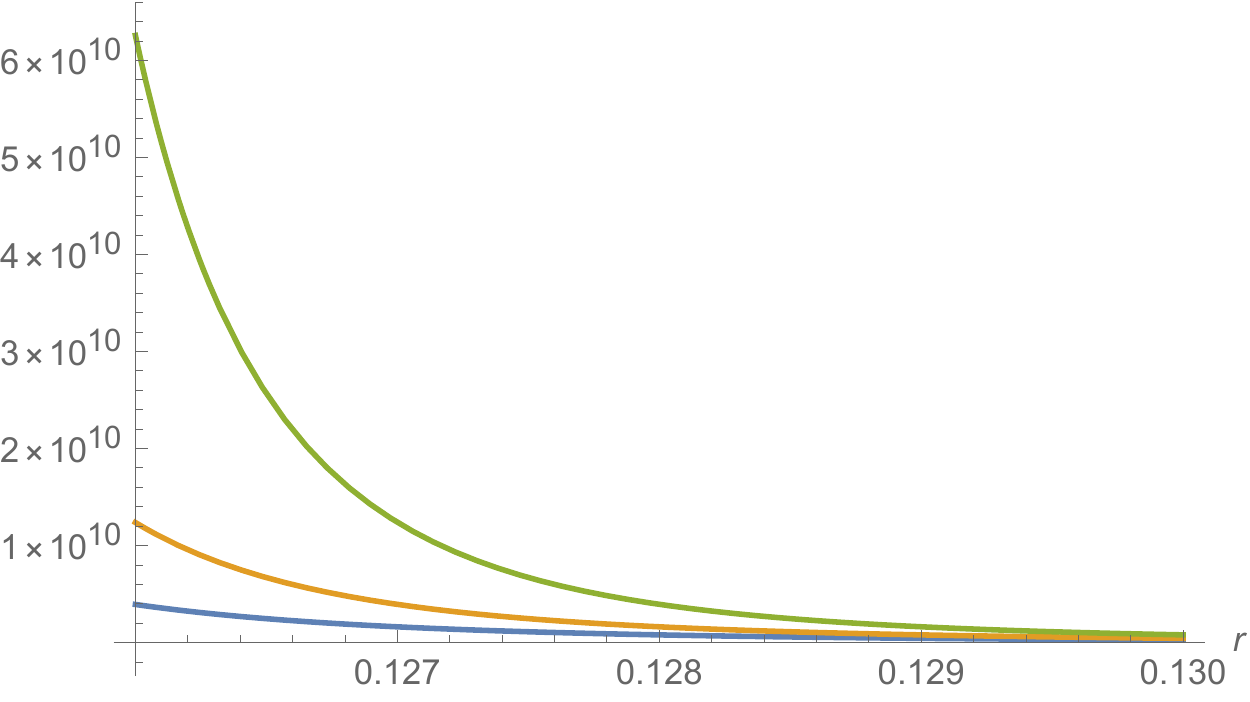}
\caption{Plot of the correlator (\ref{quattronove}),  
up to the factor $(\frac{\hbar}{4\pi})^2$, as a function of $r$ for fixed values of $r'=0.122$ (blue curve), $0.123$ (orange) and $0.124$ (green).}
\label{figsei}
\end{figure}

\begin{figure}[h]
\centering \includegraphics[angle=0, height=2.5in] {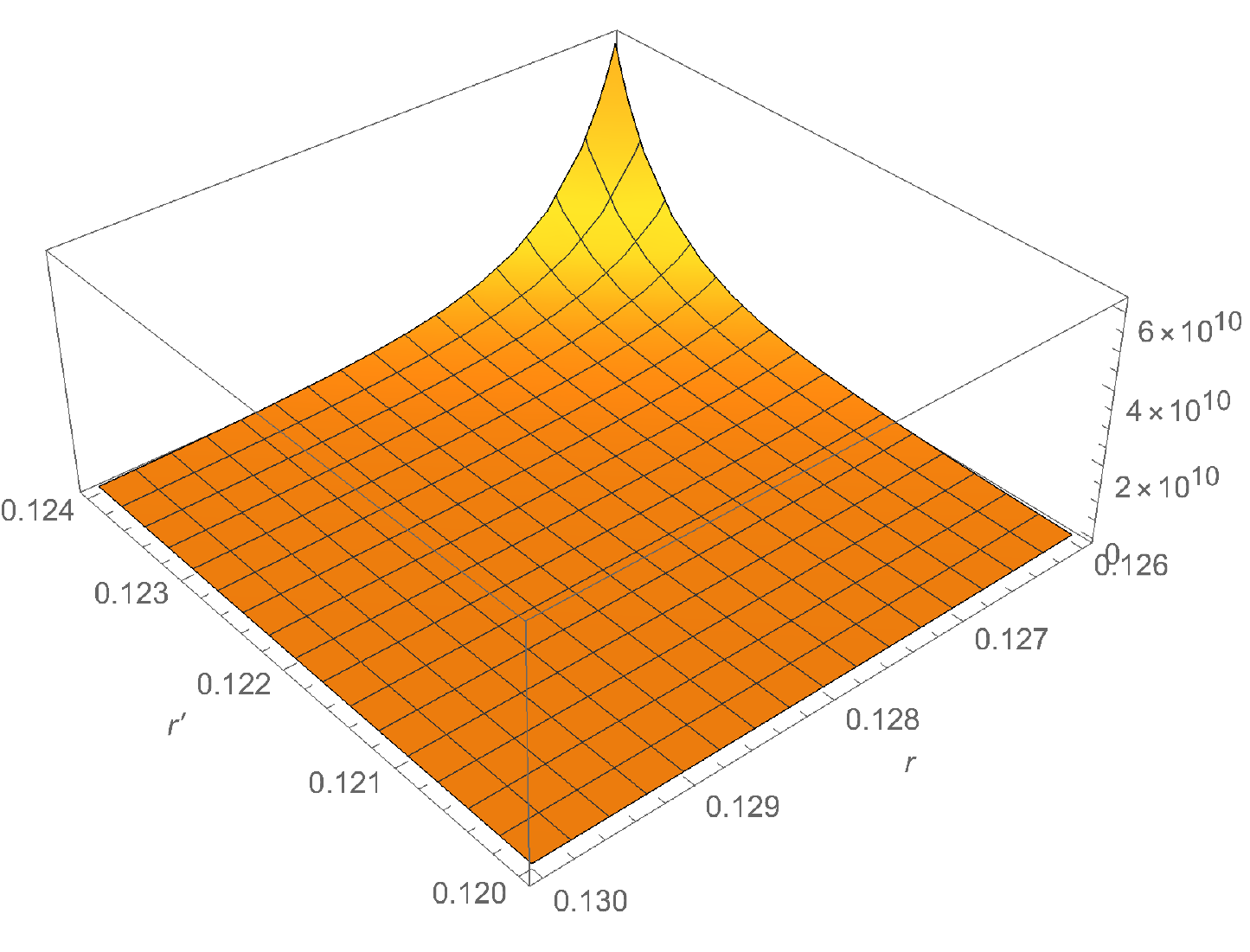}
\caption{3D Plot of the correlator (\ref{quattronove}),  
up to the factor $(\frac{\hbar}{4\pi})^2$,  for $0.120<r'<0.124$ and $0.126<r<0.130$.}
\label{figsette}
\end{figure}

From these one sees that the condition (\ref{quattrodieci}) can always be satisfied, i.e. for every $r>r_+$ a value of $r'$ exists such that $r_-<r'<r_+$ and (\ref{quattrodieci}) is satisfied. However if we plot the correlator (\ref{quattronove}) for both points close to $r_+$ no sign of the particle-partner correlation appears. In Fig. (\ref{figsei}) the correlator (\ref{quattronove}) is represented graphically at values of $r'$ fixed as a function of $r$, while in fig. (\ref{figsette}) the 3D plot is shown. 
The reason for this behaviour is that correlations only appear when when the particles and the partners emerge out of the quantum atmosphere.
\footnote{Close to the horizon the correlator is dominated by the light-cone singularity (coincidence limit in the case of equal time).} 

\begin{figure}[h]
\centering \includegraphics[angle=0, height=2in] {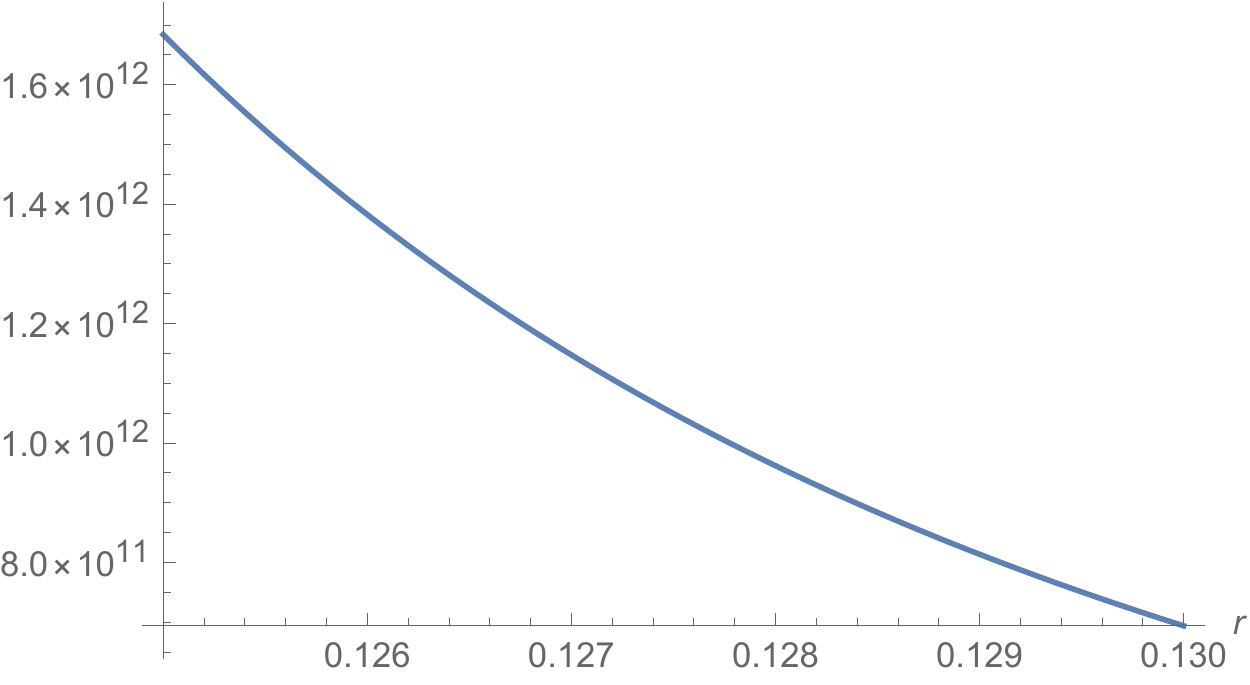}
\caption{Plot of the correlator (\ref{quattrododici}),  
up to the factor $(\frac{\hbar}{4\pi})^2$, for $t_{EF}=1$,  inner point fixed at $r'=0.0937501$ and  as a function of $r$, $0.125<r<0.130$.}
\label{figotto}
\end{figure}

\begin{figure}[h]
\centering \includegraphics[angle=0, height=2in] {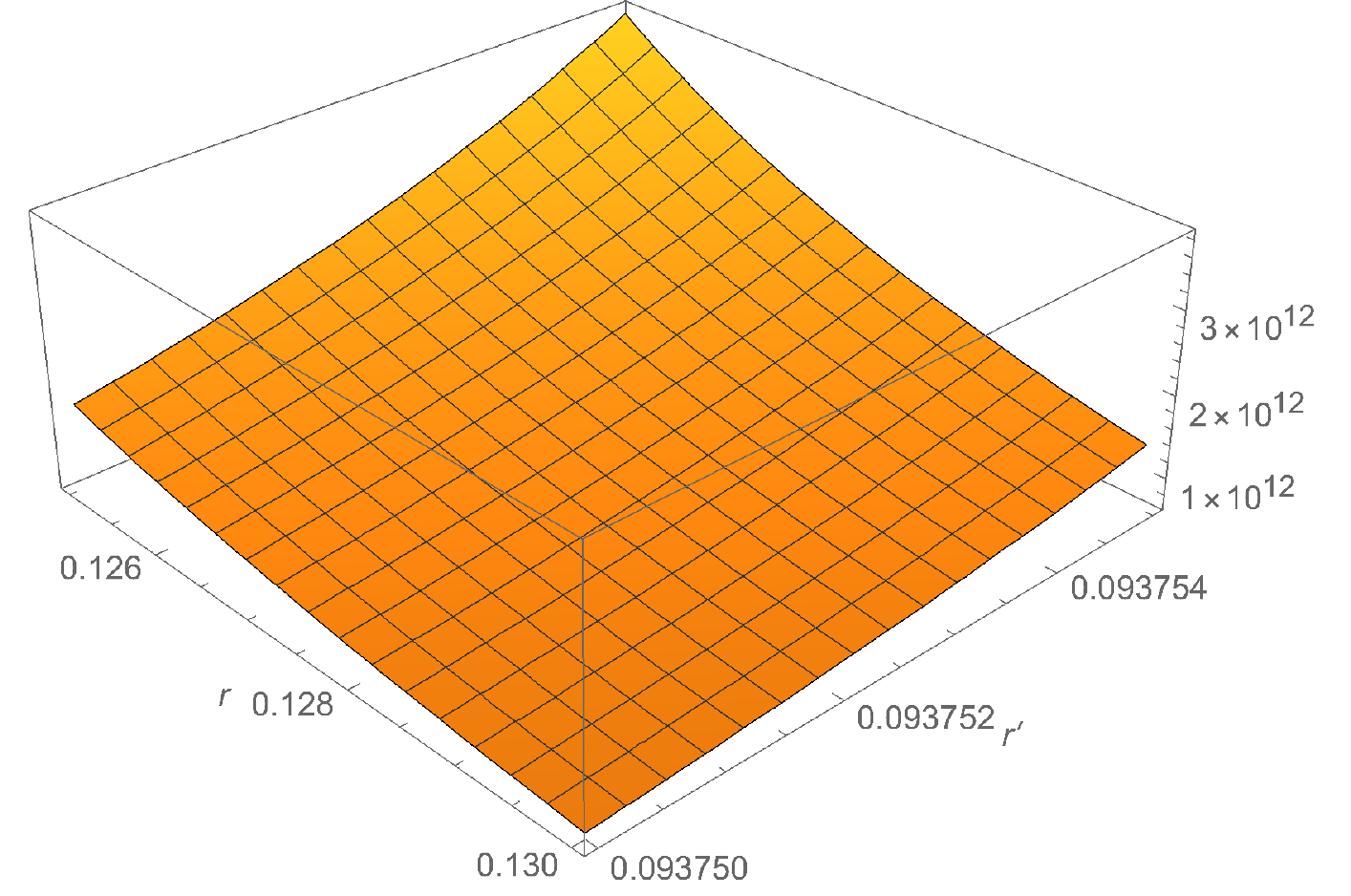}
\caption{3D Plot of the correlator (\ref{quattrododici}),   
up to the factor $(\frac{\hbar}{4\pi})^2$,  for $t_{EF}=1$ ,\ \   $0.09375<r'<0.093755$ and  $0.125<r<0.130$.}
\label{fignove}
\end{figure}

Looking at Fig. (\ref{figcinque}) one sees that all points with $r-r_+\geq \frac{1}{\kappa_+}$ are correlated with corresponding partners located very close to $r_-$. So let us consider the inner point $r'$ in this region. We have to distinguish two regimes. If the inner observer approaches the inner horizon $r_-$ at early time with respect to the formation of the BH (trajectory $A$ in Fig. (\ref{figquattro})) we have that the corresponding $u'$ approaches $-\infty$ as
\be \label{quattroundici}
u'\sim \frac{1}{\kappa_-}\ln|\kappa_-(2(r_+-r_-)-u_{in}')|\ . \ee 
If we take the other observer in the region $u\to +\infty$, using (\ref{quattroundici}) and (\ref{quattrootto}) the correlator is given by
\bea \label{quattrododici} 
&& \frac{ \langle in|\hat T_{uu}(x)\hat T_{u'u'}(x')|in\rangle}{\left(1-\sqrt{\frac{2m(r)}{r}} \right)^2\left(1-\sqrt{\frac{2m(r')}{r}} \right)^2 }
\sim  \frac{1}{ \left(1-\sqrt{\frac{2m(r)}{r}} \right)^2\left(1-\sqrt{\frac{2m(r')}{r}} \right)^2}\times  \\ &&   \frac{(\hbar \kappa_+^2\kappa_-^2)^2}{\left(4\pi \left( 2\kappa_+\kappa_-(r_+-r_-)e^{\frac{\kappa_+u-\kappa_-u'}{2}}-\kappa_+e^{\frac{\kappa_+u+\kappa_-u'}{2}}+\kappa_-e^{-\frac{\kappa_+u+\kappa_-u'}{2}}\right)^2\right)^2} \ , \nonumber \eea
which has a more complicated structure than the one in eq. (\ref{quattronove}). 
Note, using (\ref{duequindicib}), that the correlator (\ref{quattrododici}), unlike (\ref{quattronove}), is time dependent.
We represent graphically this correlator fixing the inner point very close to $r_-$ in Fig. (\ref{figotto}) and the 3D plot is shown in Fig. (\ref{fignove}). 
Again no sign of the correlations appear. The reason for this is clear from Fig. (\ref{figquattro}). 

\begin{figure}[h]
\centering \includegraphics[angle=0, height=2in] {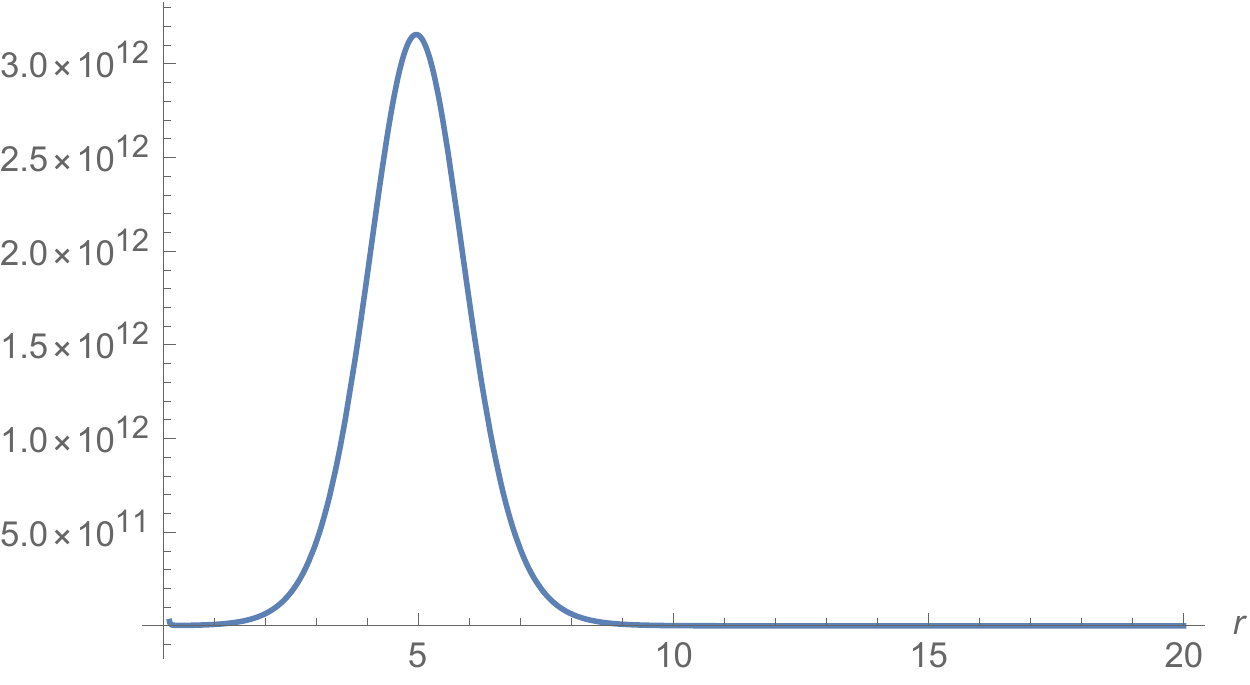}
\caption{Plot of the correlator (\ref{quattronove}),  
up to the factor $(\frac{\hbar}{4\pi})^2$, with the inner point fixed at $r'= 0.0937501$ and as a function of $r$,  $0.125<r<20$.}
\label{figdieci}
\end{figure}

\begin{figure}[h]
\centering \includegraphics[angle=0, height=2.5in] {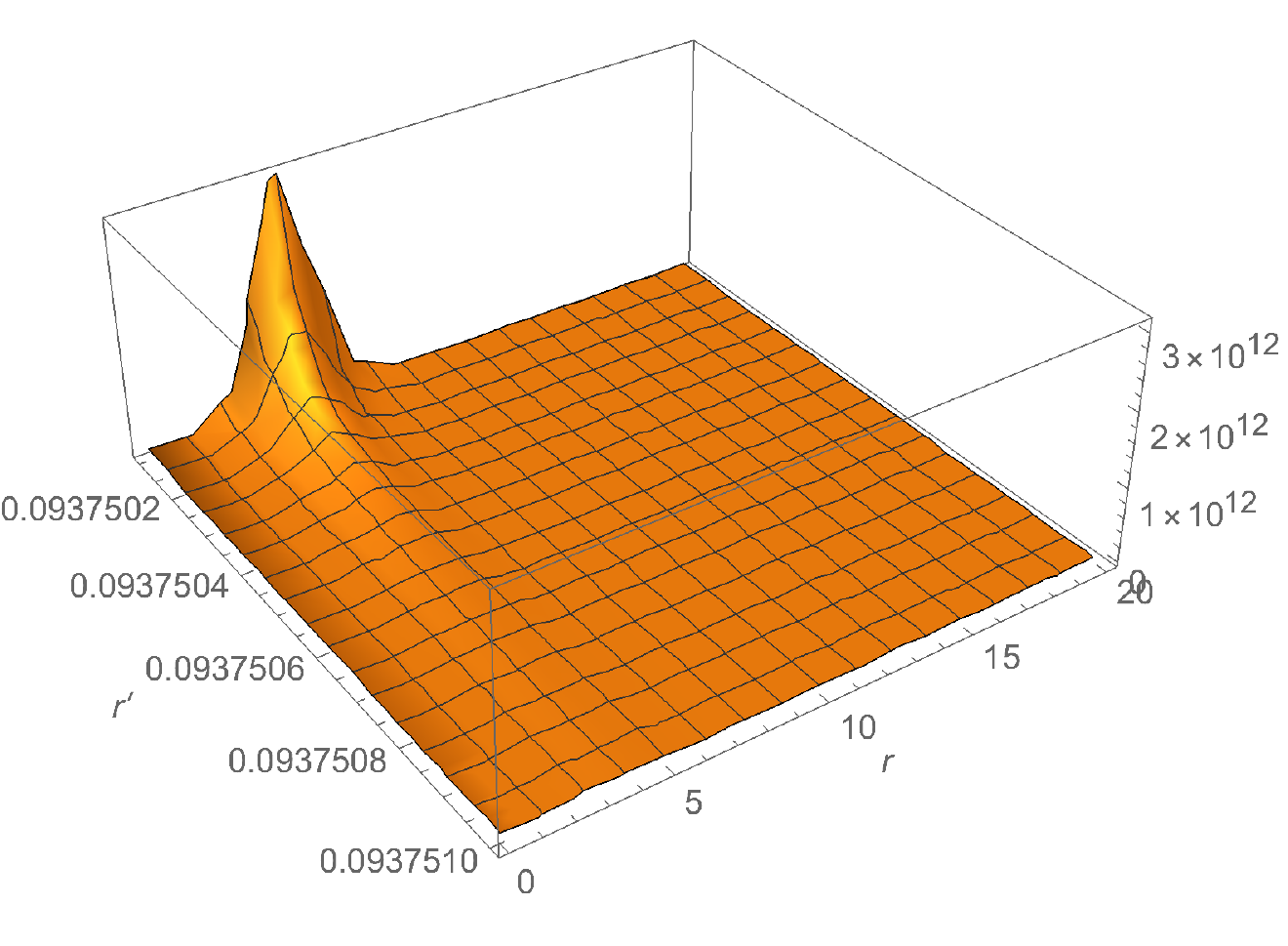}
\caption{3D Plot of the correlator (\ref{quattronove}),  
up to the factor $(\frac{\hbar}{4\pi})^2$, for  $ 0.0937501< r'< 0.093751 $ and  $0.125<r<20$.}
\label{figundici}
\end{figure}

\noindent The partners of the Hawking particles pile up at $r_-$ and only at late time they are intercepted by our inner observer close to $r_-$ (see trajectory $B$ in Fig. (\ref{figquattro})). So in order to reveal the correlation we have to consider the limit in which $u'$ is given by (\ref{quattrosette}) and $r'$ is close to $r_-$. The correlator is given by eq. (\ref{quattronove}) with $r'\to r_-$ and the corresponding plots are in Figs. (\ref{figdieci}) and (\ref{figundici}). 
We see clearly the appearance of the foreseen peak. The corresponding $r$ is not exactly the one given by the condition (\ref{quattrodieci}) because of the nontrivial role played by the geometric prefactor containing $\sqrt{ \frac{2m(r)}{r}} $. 

Completely different is the situation in which the inner observer approaches the Cauchy horizon. In that case, since this observer has $E=-1$, see eq. (\ref{quattrodue}), the $G(x,x')$ correlator  becomes  
\be \label{quattrotredici} G(x,x')= \frac{ \langle in|\hat T_{uu}(x)\hat T_{u'u'}(x')|in\rangle}{\left(1- \sqrt{ \frac{2m(r)}{r}}\right)^2\left(1+\sqrt{ \frac{2m(r')}{r'}} \right)^2}+  \frac{ \langle in|\hat T_{vv}(x)\hat T_{v'v'}(x')|in\rangle}{\left(1+ \sqrt{ \frac{2m(r)}{r}} \right)^2\left(1-\sqrt{ \frac{2m(r')}{r'}}\right)^2}\ , \ee 
and the dominant, now diverging contribution comes from the last term, namely the $\hat T_{vv}$ correlator,
\be \label{quattroquattordici} 
\frac{ \langle in|\hat T_{vv}(x)\hat T_{v'v'}(x')|in\rangle}{\left(1+ \sqrt{ \frac{2m(r)}{r}}\right)^2\left(1-\sqrt{ \frac{2m(r')}{r'}}\right)^2}
\sim \frac{1}{ \left(1+\sqrt{ \frac{2m(r)}{r}} \right)^2\left(1- \sqrt{ \frac{2m(r')}{r'}} \right)^2}\left( \frac{\hbar }{4\pi (v-v')}\right)^2 \  \ee
showing explicitly the divergence as $r'\to r_-$ due to the vacuum polarization. 

\section{Conclusions}
\label{s5}

In this paper, within the framework of Quantum Field Theory in curved spacetime, we studied, in a two dimensional section of the Reissner-Nordstr\"om BH
spacetime formed by the collapse of a null shell, the expectation values of the energy momentum tensor operator for a massless scalar field showing how vacuum polarization and particles creation mix up and one can not disentangle the two; both contribute to the measured energy density.
Only sufficiently far away from the horizon, the tensor describes an outgoing flux of particles at the Hawking temperature. To reveal the existing correlations between Hawking particles and their partners one has to look at correlation functions.

In a Schwarzschild BH studying equal time correlators we found \cite{Balbinot:2021bnp} that the above correlations do not show up because the partner is swallowed by the central singularity before the correlated Hawking particle emerges from the quantum atmosphere. To reveal the correlations one has to consider unequal time correlators in order to catch the partner before it disappears into the singularity.

In a Reissner-Nordstr\"om BH the picture changes significantly. The spacetime structure inside the event horizon is characterised by the presence of an inner horizon before reaching the singularity. At this horizon all the partners pile up asymptotically never reaching the singularity. So equal time correlators  across the horizon show up, at late time, a significant enhancement when the inner point is taken close to the inner horizon singling out the quantum entanglement of the Hawking particles and the partners.
On the ingoing part of the inner horizon (the Cauchy horizon), on the other hand, we found that the correlators diverge like the energy momentum tensor, the divergence being caused by infinite vacuum polarization there.

We should stress that our analysis is restricted to the gravitational case discussed within the framework of Einstein's General Relativity Theory.
In acoustic BH realised with BECs, because of the modified dispersion relation which allows for "superluminally" propagating modes, the partners do not pile up at the inner horizon, but bounce back and forth between the inner and the event horizon producing the so called laser effect \cite{Corley:1998rk} which, in turn,  causes the 
rise of a dynamical instability \cite{Coutant:2009cu}.  

\acknowledgments
A.F. acknowledges partial financial support by the Spanish Ministerio de Ciencia e Innovaci\'on Grant No. PID2020–116567 GB-C21 funded
by Grant No. MCIN/AEI/10.13039/501100011033, and
the Project No. PROMETEO/2020/079 (Generalitat
Valenciana).




\end{document}